\documentclass[preprint,aps,nofootinbib]{revtex4}
\usepackage{graphicx}
\usepackage{amsmath}
\usepackage{amsfonts}
\usepackage{amssymb}
\usepackage{color}%
\providecommand{\U}[1]{\protect\rule{.1in}{.1in}}

\newcommand{\f}{\begin{equation}}
\newcommand{\ff}{\end{equation}}
\newcommand{\fa}{\begin{eqnarray}}
\newcommand{\ffa}{\end{eqnarray}}

\begin{document}
\title{A note on entropic force and brane cosmology}
\author{Yi Ling}
\email{yling@ncu.edu.cn} \affiliation{Center for Relativistic Astrophysics and High Energy Physics,Department of Physics, Nanchang University, 330031, China}
\author{Jian-Pin Wu}
\email{jianpinwu@yahoo.com.cn} \affiliation{Department of Physics, Beijing Normal University, Beijing 100875, China}

\begin{abstract}

Recently Verlinde proposed that gravity is an entropic force
caused by information changes when a material body moves away from
the holographic screen. In this note we apply this argument to
brane cosmology, and show that the cosmological equation can be
derived from this holographic scenario.

\end{abstract}
\maketitle

%\section{Introduction}

Recently Verlinde argued that gravity may be an entropic force
caused by information changes when a material body moves away from
the holographic screen\cite{Verlinde:2010hp}. Quantitatively, when
a test particle or excitation moves apart from the holographic
screen, the magnitude of the entropic force on this body has the
form
\begin{equation} \label{ef}
F\triangle x=T \triangle S,
\end{equation}
where $\triangle x$ is the displacement of the particle from the
holographic screen, while $T$ and $\triangle S$ are the
temperature and the entropy change on the screen, respectively.
Verlinde further postulates that the entropy change is linear in
the displacement, which is motivated by Bekenstein's argument,
\begin{equation} \label{ec}
\triangle S=2\pi k_B\frac{mc}{\hbar}\triangle x,
\end{equation}
where $m$ is the mass of the test particle. Then the entropic
force is simplified as
\begin{equation} \label{ef2}
F=2\pi k_BT\frac{mc}{\hbar},
\end{equation}
implying that the magnitude of the force depends only on the
temperature of the holographic screen and the mass of the test
particle. Remarkably, employing the holographic principle and the
equipartition rule of energy
\begin{equation} \label{bits}
N=\frac{Ac^{3}}{G\hbar},\ \ \ \ \ \ Mc^2={1\over 2}Nk_BT,
\end{equation}
where $A$ is the area of the holographic surface, one finds that
Equation (\ref{ef2}) is nothing but the Newton's law of
gravitation!
\begin{equation} \label{bits}
F=G{Mm\over R^2}.
\end{equation}
Similar consideration can also be found in
\cite{Padmanabhan:2003pk,Padmanabhan:2009vy}. Furthermore,
Verlinde shows that the law of gravitation can be extended to the
relativistic case such that the Einstein equations can be
constructed closely following the thermodynamical description of
general relativity originally advocated by
Jacobson\cite{Jacobson:1995ab}. The key point there is to connect
the energy momentum $T_{ab}$ with the spacetime curvature with the
use of the Tolman-Komar's definition of active gravitational mass
contained inside an arbitrary volume. In \cite{Padmanabhan:2010qr}, 
the Friedmann-Robertson-Walker (FRW) universe has been explicitly discussed. 
Furthermore, \cite{Cai:2010hk} has also demonstrated explicitly that the Friemdann equations
for FRW universe can be derived following this route with the help of holographic principle and
the equipartition rule of energy. In this short note we intend to
argue that the cosmological equation for a brane world can be
derived in a parallel manner. Stimulated by the remarkable work by
Verlinde, very recent relevant work on entropic force conjecture
can be found in
\cite{Shu:2010nv,Smolin:2010kk,Li:2010cj,Zhang:2010hi,Gao:2010fw,Wang:2010jm,Wei:2010sw,Caravelli:2010F}.

For simplicity we consider a spatially flat $FRW$ universe without
cosmological constant embedded in a Randall-Sundrum model. Its
cosmological equation can be derived from the bulk action and for
details we refer to \cite{Maartens:2003tw}. The main difference
from the standard Friedmann equation in $GR$ is that the brane
equation receives a $\rho^2$ correction term. Next we intend to
reconstruct this equation following the strategy in
\cite{Verlinde:2010hp}. Consider a compact spatial region with a
physical radius $R=ar$, where $a$ is the scale factor and $r$ is
the radial comoving coordinate. Because of the spherical symmetry
its boundary is an equipotential surface such that we may treat it
as a holographic screen. Then from holographic principle the
number of bits on this screen can be specified as
\begin{equation} \label{bits}
N=\frac{Ac^{3}}{G\hbar}.
\end{equation}
Furthermore, since a comoving observer at $r$ feels a radial
acceleration $a_{r}=-\ddot{a}r$,
%\begin{equation} \label{acceleration}
%a_{r}=-\ddot{a}r~,
%\end{equation}
then following Unruh's argument we may define a temperature on the
holographic screen to be
\begin{equation} \label{Unruh}
\mathcal{T}=\frac{1}{2\pi k_{B}c}\hbar a_{r}~.
\end{equation}

Through the equipartition rule, the total energy on screen is
related to the temperature as
\begin{equation} \label{equipartition}
E=\frac{1}{2}Nk_{B}\mathcal{T}~.
\end{equation}

All above quantities describe the properties of spacetime
geometry. Next we need connect these quantities with the
distribution of matter sources. As assumed in
\cite{Verlinde:2010hp}, a matter source with active gravitational
mass $\mathcal{M}$ which drives a test particle accelerating would
emerge inside the compact region surrounded by the holographic
screen and its mass satisfies the relation
\begin{equation} \label{EMC}
E=\mathcal{M}c^{2}~.
\end{equation}
Now, to derive the cosmological equations the key point is
expressing the mass $\mathcal{M}$ in terms of $T_{ab}$ from the
side of matter. In brane scenario, we find that the induced active
gravitational mass on the brane has the form
\begin{equation} \label{activemassb}
\mathcal{M}=2\int_{\mathcal{V}}dV[(T_{\mu\nu}+\frac{6}{\lambda}S_{\mu\nu})-\frac{1}{2}(T+\frac{6}{\lambda}S)g_{\mu\nu}]u^{\mu}u^{\nu}~,
\end{equation}
where $T_{\mu\nu}$ is the energy tensor of the matter, and
$\lambda$ is the brane tension which is tuned with the five
dimensional cosmological constant through $\lambda\sim
\frac{-\Lambda_5}{8\pi G_4}$. While $S_{\mu\nu}$ is a high energy
correction term carrying local bulk effects on the brane. In terms
of $T_{ab}$, $S_{\mu\nu}$ can be expressed as
\begin{equation} \label{highenergyc}
S_{\mu\nu}=\frac{1}{12}TT_{\mu\nu}-\frac{1}{4}T_{\mu\alpha}T^{\alpha}_{\nu}+\frac{1}{24}(3T_{\alpha\beta}T^{\alpha\beta}-T^{2})g_{\mu\nu}~.
\end{equation}

For simplicity we consider a perfect fluid with the stress-energy
tensor
\begin{equation} \label{pfluidtensor}
T_{\mu\nu}=(\rho+p)u_{\mu}u_{\nu}+pg_{\mu\nu}~.
\end{equation}
Its continuity equation reads as
\begin{equation} \label{continuitye}
\dot{\rho}+3H(\rho+p)=0~.
\end{equation}

Then the stress-energy scalar can be expressed as
\begin{equation} \label{pfluidscalar}
T=-\rho+3p~.
\end{equation}

Therefore, the high energy correction term can be obtained
explicitly
\begin{equation} \label{highenergyc1}
S_{\mu\nu}=\frac{1}{6}(\rho^{2}+\rho p)u_{\mu}u_{\nu}+\frac{1}{12}(\rho^{2}+2\rho p)g_{\mu\nu}~.
\end{equation}
\begin{equation} \label{highenergyc2}
S=\frac{1}{6}\rho^{2}+\frac{1}{2}\rho p~.
\end{equation}

Substituting these results into equation (\ref{activemassb}), we
find the active gravitational mass in brane cosmology can be
expressed as

\begin{equation} \label{activemassb1}
\mathcal{M}=\int_{\mathcal{V}}dV(\rho+3p+\frac{2\rho^{2}}{\lambda}+\frac{3\rho p}{\lambda})=\frac{4\pi}{3}(\rho+3p+\frac{2\rho^{2}}{\lambda}+\frac{3\rho p}{\lambda})a^{3}r^{3}~,
\end{equation}

Combining eqs. (\ref{Unruh}), (\ref{bits}),
(\ref{equipartition}), (\ref{EMC}) and (\ref{activemassb1}) , we can
derive the following equation
\begin{equation} \label{BCE1}
\frac{\ddot{a}}{a}=-\frac{4\pi G}{3}(\rho+3p+\frac{2\rho^{2}}{\lambda}+\frac{3\rho p}{\lambda})~.
\end{equation}
Multiplying $\dot{a}a$ on both sides of eq. (\ref{BCE1}), and with
the help of the continuity equation (\ref{continuitye}), we
integrate the resulting equation and yield
\begin{equation} \label{BCE2}
H^{2}+\frac{k}{a^{2}}=\frac{8\pi
G}{3}\rho(1+\frac{\rho}{2\lambda})~.
\end{equation}

Therefore, we conclude that the Friedmann equations in brane
cosmology can be derived following the proposal that gravity is an
entropic force.

\section*{Acknowledgement}
Y. Ling is partly supported by NSFC(No.10875057), Fok Ying Tung
Education Foundation(No. 111008), the key project of Chinese
Ministry of Education(No.208072) and Jiangxi young
scientists(JingGang Star) program. He also acknowledges the
support by the Program for Innovative Research Team of Nanchang
University. J. P. Wu is partly supported by NSFC(No.10975017).

\end{document}